\documentclass[runningheads]{llncs}
\usepackage[T1]{fontenc}
\usepackage{graphicx}
\usepackage{bm}
\usepackage{color}
\usepackage{amsmath}
\usepackage[misc,geometry]{ifsym} 
\usepackage{amssymb}
\usepackage{amsmath}
\usepackage{enumitem}
\usepackage{multirow}
\usepackage{booktabs}
\usepackage{adjustbox}
\usepackage{subfigure}
\usepackage{textcase}
\usepackage{float}

\newcommand{\eg}{\textit{e}.\textit{g}.}

\def\model{FedMSA}
\def\modelall{FedSAM}
\begin{document}
\title{FedFMS: Exploring Federated Foundation Models for Medical Image Segmentation}
\titlerunning{FedFMS}
\author{Yuxi Liu\inst{1} \and
Guibo Luo\inst{1}\textsuperscript{(\Letter)} \and
Yuesheng Zhu\inst{1}\textsuperscript{(\Letter)}
}
%1{Liu, Yuxi}
%2{Luo, Guibo}
%3{Zhu, Yuesheng}

\authorrunning{Yuxi Liu et al.}
\institute{School of Electronic and Computer Engineering, Peking University, Shenzhen, China\\
\textsuperscript{\Letter}Correspondences: \email{\{luogb,zhuys\}@pku.edu.cn}
%\url{http://www.springer.com/gp/computer-science/lncs}
}
\maketitle              % typeset the header of the contribution
\begin{abstract}
Medical image segmentation is crucial for clinical diagnosis. The Segmentation Anything Model (SAM) serves as a powerful foundation model for visual segmentation and can be adapted for medical image segmentation. However, medical imaging data typically contain privacy-sensitive information, making it challenging to train foundation models with centralized storage and sharing. To date, there are few foundation models tailored for medical image deployment within the federated learning framework, and the segmentation performance, as well as the efficiency of communication and training, remain unexplored. In response to these issues, we developed Federated Foundation models for Medical image Segmentation (FedFMS), which includes the Federated SAM (FedSAM) and a communication and training-efficient Federated SAM with Medical SAM Adapter (FedMSA). Comprehensive experiments on diverse datasets are conducted to investigate the performance disparities between centralized training and federated learning across various configurations of FedFMS. The experiments revealed that FedFMS could achieve performance comparable to models trained via centralized training methods while maintaining privacy. Furthermore, FedMSA demonstrated the potential to enhance communication and training efficiency. Our model implementation codes are available at https://github.com/LIU-YUXI/FedFMS. 
\keywords{Medical image segmentation \and Federated learning \and Foundation model.}
\end{abstract}
\section{Introduction}
\label{sec:intro}
Medical image segmentation aims to identify and separate structures or regions in medical images~\cite{wang2022medical,norouzi2014medical}, which is crucial for clinical care. Very recently, the Segmentation Anything Model (SAM)~\cite{kirillov2023segmentanything} has garnered widespread attention as a powerful foundation model for visual segmentation. Many works fine-tuning SAM in the medical images have achieved advanced results, such as the Medical SAM Adapter (MSA)~\cite{wu2023msa}, 3DSAM-adapter~\cite{gong20233dsam} and~\cite{xie2024sam}.

However, medical imaging data typically contain privacy-sensitive information, making it difficult to centralized storage and sharing~\cite{li2019privacy,sohan2023systematic}. Moreover, training large models often involves data that is distributed across various geographic locations or institutes. Transmitting large volumes of data can lead to increased communication costs and delays in transmission. Federated learning~\cite{konevcny2016federated} offers a solution by enabling model training on distributed datasets without the need to centralize data in one location~\cite{liu2021feddg}. Additionally, distributed training of large models allows for the distribution of computational requirements~\cite{rauniyar2023federated}. 

To date, deploying foundation models for medical images within the federated learning framework is rare. There are two main issues that remain unexplored: First, can foundation models trained based on federated learning harness the powerful capabilities of foundation models, and maintain performance comparable to those trained based on centralized training when facing Non-Independent and Identically Distributed (Non-IID) datasets? Second, the federated learning of foundation models requires significant communication resources and training costs, is there a more efficient method for its federated learning training?

To address the above issues, we have collected a large number of real multi-center medical datasets and developed \underline{Fed}erated \underline{F}oundation models for \underline{M}edical image \underline{S}egmentation (\textbf{FedFMS}) to investigate both its performance of segmentation and training efficiency. FedFMS includes two federated foundation models, the Federated SAM (FedSAM) and a communication and training-efficient Federated SAM with Medical SAM Adapter (FedMSA). 
For FedSAM, we fine-tune all parameters of the pre-trained SAM on each client. For FedMSA, we efficiently fine-tune the parameters of the adapters and decoder of the pre-trained MSA on each client. Then, we aggregate the parameters on the global server using the FedAvg~\cite{mcmahan2017fedavg} algorithm. To our knowledge, this study is the first comprehensive investigation into the application of federated foundation models within the medical domain. Our contributions can be summarized as follows:

(1) \textbf{Dataset Collection}. We have collected various multi-institutional datasets to serve as benchmarks for evaluating the performance of federated foundation models in medical image segmentation. This offers comprehensive and reliable evaluation data for federated medical segmentation.

(2) \textbf{Model Development}. We have developed a federated learning framework named FedSAM based on the foundation model SAM, which enables distributed training of medical images and demonstrates stability and effectiveness. To further explore more efficient methods, we have also built the FedMSA framework. These models could serve as baselines and be beneficial for further promoting the federated foundation models for medical image segmentation.

(3) \textbf{Experimental Analysis}. We have conducted an in-depth investigation into the performance disparities between centralized training and federated learning across different configurations of FedFMS with various datasets. Our investigation will provide an insight overview of the feasibility and effectiveness of a federated large-scale model for medical images in real-world clinical settings.

\section{Method}
\label{sec:method}
\subsection{Preliminary Methods}
\subsubsection{SAM Architecture}
SAM is a large data-driven image segmentation model. It constructs a dataset named SA-1B, consisting of 11 million images and 100 million masks, to drive its training. %SAM comprises three main components: an image encoder, a prompt encoder, and a mask decoder.
The image encoder utilizes a standard Vision Transformer (ViT)~\cite{dosovitskiy2020vit1,steiner2021vit2} pre-trained by Masked Autoencoders.  In our study, we set SAM using the ViT-B/16 and ViT-L/16 variants. ViT-B/16 represents the base-scale version of ViT, implemented with 768 convolutional kernels. ViT-L represents the large-scale version, implemented with 1024 convolutional kernels. The output of the image encoder is a 16× downsampled embedding of the input image.
% SAM's prompt encoder can be either sparse (points, boxes) or dense (masks).
The mask decoder is a lightweight modified Transformer decoder block that includes bidirectional cross-attention and a dynamic mask prediction head. % for predicting the target mask.
% In our work, to achieve simultaneous segmentation of multiple types (such as segmentation of the cup and disc in fundus images), the input prompts and corresponding prompt encoders are omitted. We directly adopt an encoding-decoding approach from SAM to obtain segmentation results, as shown in Figure \ref{fig:fedsam} (b).
% The decoder employs bidirectional cross-attention to learn the interaction between prompts and image embeddings. % . The decoder employs bidirectional cross-attention to learn the interaction between prompts and image embeddings. Subsequently, SAM upsamples the image embeddings, and a multilayer perceptron (MLP) maps the output tokens to a dynamic linear classifier

\subsubsection{MSA Architecture}
MSA efficiently fine-tunes the SAM architecture for medical images to enhance its medical capabilities. Fine-tuning allows the model to retain the knowledge gained from extensive data while strengthening its abilities in new domains. In the encoder, MSA freezes the pre-trained SAM parameters and inserts two adapter modules at each ViT block. The adapter is a bottleneck model that sequentially uses down-projection, ReLU activation, and up-projection. Both projections are implemented using simple MLP layers. 
MSA's decoder is the same as SAM's. MSA fine-tunes all parameters of the decoder.

\subsection{Federated Foundation Models}
\begin{figure}[t]
\centering
\includegraphics[width=\linewidth]{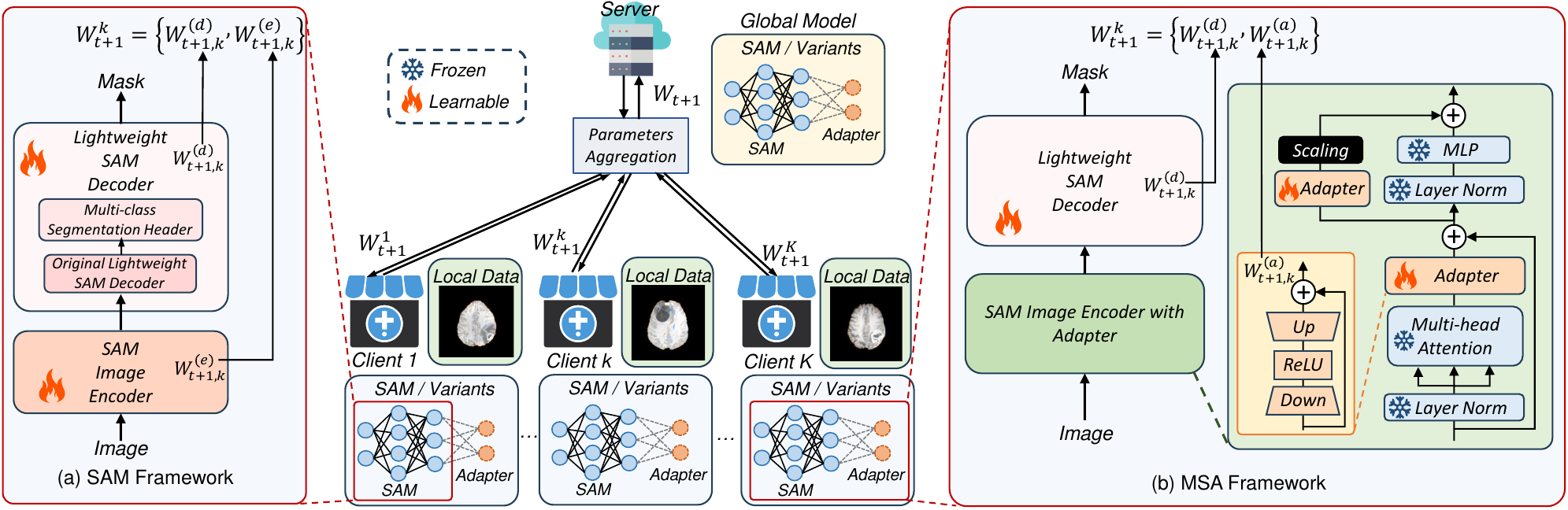}
%\vspace{-0.1in}
\caption{The illustration of \modelall\ and \model. \modelall\ is a federated learning framework with (a) SAM. \model\ is a federated learning framework with (b) MSA.}
% \vspace{-0.1in}
\label{fig:fedsam}
\end{figure}
Figure \ref{fig:fedsam} illustrates the FedFMS framework, comprising multiple clients for local training and a server for parameter aggregation, all utilizing the same foundational model (\eg, SAM or MSA). The federated learning of SAM and its more efficient variant MSA corresponds to FedSAM and FedMSA, respectively.
% Figure \ref{fig:fedsam} illustrates the FedSAM framework and its variant FedMSA.%, which is implemented with the FedAvg algorithm.% In this section, we introduce the client-side model training process and the server-side model aggregation process.
\subsubsection{Client-side Model Training}
Each client possesses a fixed local dataset and sufficient computational resources to perform mini-batch updates. The number of clients is $K$. Each client adopts the same BCE loss and the same model (SAM or MSA), which is initialized with pre-trained SAM parameters before training. %To simulate real-world scenarios and facilitate the model as a benchmark for domain shift,   The datasets for each client are sourced from different origins.

\noindent{\textbf{FedSAM}} To achieve simultaneous segmentation of multiple classes, we omit the input prompts and prompt encoder, perform a multi-class segmentation header by adopting a two-dimensional convolution with a $1 \times 1$ kernel after the original SAM decoder, mapping the output mask to $H \times W \times c$, where $c$ is the number of segmentation classes, $H$ is the height and $W$ is the width of the predictive mask. Our SAM is shown in Figure \ref{fig:fedsam} (a). The local SAM is initialized by global parameters $W_t=\{W^{(d)}_t,W^{(e)}_t\}$, where $W^{(e)}_t$ is the parameters of the encoder and $W^{(d)}_t$ is the parameters of the decoder. In the $k$-th client, the updated parameters is $W_{t+1}^{k}=\{W^{(d)}_{t+1,k},W^{(e)}_{t+1,k}\}$, where $W^{(e)}_{t+1,k}$ is the updated parameters of the encoder and $W^{(d)}_{t+1,k}$ is the updated parameters of the decoder.% (such as segmentation of the cup and disc in fundus images)
%We adopt an encoding-decoding approach from SAM to obtain segmentation results, as shown in Figure \ref{fig:fedsam} (a).

\noindent{\bf{FedMSA}}
For MSA, we fine-tune the parameters of adapters in the encoder (denoted as $W^{(a)}_t$), and all parameters in the decoder. 
Our MSA's decoder adopts the same multi-class segmentation decoder as our SAM.
The features obtained by fine-tuning the encoder propagate to the top layers of the decoder, so all parameters of the decoder need to be fine-tuned. The parameter of the SAM decoder is lightweight, resulting in a low fine-tuning cost. The structure of MSA is shown in Figure \ref{fig:fedsam} (b). We use MSA to build a more efficient federated learning framework for three reasons. (1) MSA performs well in fine-tuning tasks for medical image segmentation. (2) Only training adapters and decoder requires less computational cost compared to training the entire SAM. (3) During global parameter aggregation, only the parameters of the adapters and decoder need to be transmitted and calculated. During the federated $t$-th round, the local model is initialized by fetching global model parameters $W_t=\{W^{(d)}_t,W^{(a)}_t\}$ from the server. In the $k$-th client, the updated parameters is $W_{t+1}^{k}=\{W^{(d)}_{t+1,k},W^{(a)}_{t+1,k}\}$, where $W^{(a)}_{t+1,k}$ is the updated parameters of the adapters.% and $W^{(d)}_{t+1,k}$ is the parameters of decoder in client $k$.

%After local training iterations, the new parameters $W_{t+1}^{k}$ of the model participate in the aggregation shared with the server.
\subsubsection{Server-side Model Aggregation}
The server distributes a global model and receives synchronized updates from all clients at each federated round. 
We use FedAvg as the aggregation method.
The aggregation is formalized as $W_{t+1} \leftarrow \frac{1}{\sum_{k=1}^{K}N_k^{(local)}}\sum_{k=1}^{K}(N_k^{(local)}\cdot W^k_{t+1})$, where $N_k^{(local)}$ is the amount of data in client $k$. 
Though more complex algorithms could also be considered, FedAvg has shown good performance due to the strong generalization capabilities of SAM.
% In fact, during client training, we only fine-tune a portion of the parameters of SAM and the adapter. During aggregation, only the unfrozen parameters need to be communicated and weighted, which helps to save communication and computation costs. During aggregation, each client's parameters are weighted based on the amount of data, denoted as $N_k^{(local)}$. The parameters of the global model in the $t$-th round are defined as $W_t$. 
\subsection{Dataset Preparation}
\begin{figure}[t]
\centering
%\vspace{-0.05in}
\includegraphics[width=1\linewidth]{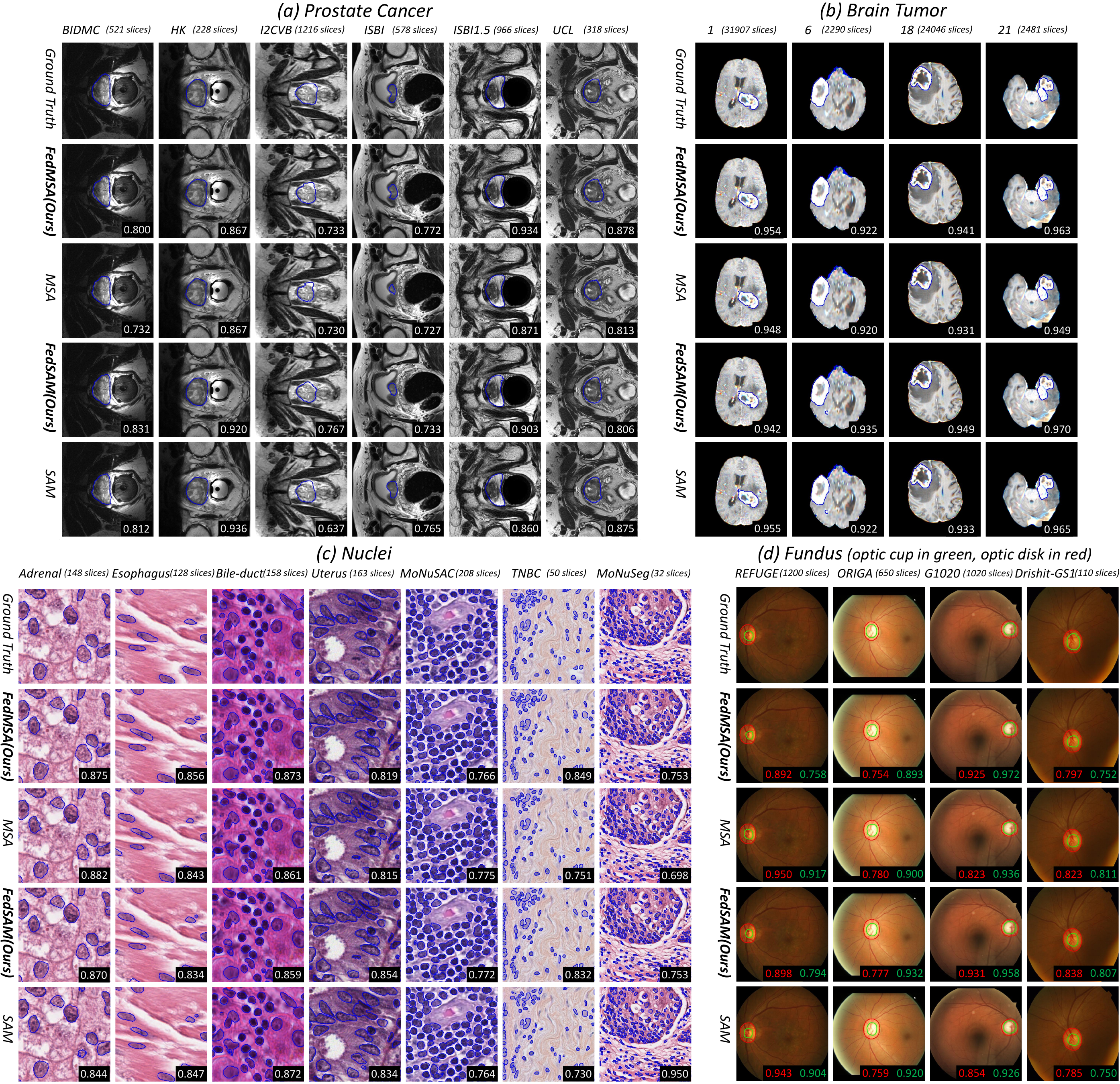}
%\vspace{-0.05in}
\caption{The example of images and Ground Truth from different clients across four datasets (a-d) and the comparison in the results of \model, MSA, \modelall~and SAM. The bottom right corner of each image indicates the Dice for it or the whole nii.}
%\caption{The display of images and segmentation labels (Groundtruth) from different clients (A, B, C, ...) across four federated learning datasets (a-d) and the comparison in the results of \model, MSA, \modelall~and SAM on these datasets. The bottom right corner of each image indicates its Dice based on the segmentation result.}
%\vspace{-0.05in}
\label{fig:result}
\end{figure}

We collected and constructed four Non-IID federated learning datasets with different modalities and types from public datasets. The example cases and sample numbers of each data are presented in Figure \ref{fig:result}. Following the methodology of MSA, all images are preprocessed to a shape of $1024 \times 1024 \times 3$ before input, and the size of the output mask is $256 \times 256$.% We build these non-IID federated learning datasets to train and test the \model\ and \modelall, which enables \model\ and \modelall\ to serve as a benchmark model for experimental comparisons in a wider range of domain generalization tasks. %In real-world scenarios, data distributions and features may vary between different local devices or users. 

% \noindent\textbf{Non-IID Datasets.}\vspace{-0.05in}
\begin{itemize}[leftmargin=*]
\item \textbf{Prostate Cancer}. Extracted from public prostate cancer MRI imaging datasets from various medical institutions \cite{litjens2014evaluation,lemaitre2015computer} and NCI-ISBI 2013.
\item \textbf{Brain Tumor}. Derived from FeTS2022~\cite{pati2021federated}, which is a collection of multi-institutional clinical acquisition mp-MRI scans of gliomas. 
The segmentation target we use is the GD-enhancing tumor (ET - label 4) on T1ce images.
\item \textbf{Nuclei}. It is a nuclei segmentation dataset from \cite{gamper2020pannuke,gamper2019pannuke,verma2021monusac2020,naylor2018segmentation,kumar2017dataset}. Cells from different tissues in the PanNuke dataset are distributed across different clients. %For non-IID purposes, 
\item \textbf{Fundus}. Gathered from four distinct fundus photography images datasets \cite{sivaswamy2015comprehensive,orlando2020refuge,bajwa2020g1020,zhang2010origa} for optic cup (OC) and optic disc (OD) segmentation tasks.
\end{itemize}

Brain Tumor and Prostate Cancer images are in 3D nii format, while SAM can only handle 2D images. Therefore, we slice them along the depth direction, converting a 3D image with depth $d$ into $d$ slices of 2D images. Since the maximum pixel value in nii format is much larger than 255, we calculate the 1st percentile of pixel values for each nii file as the upper bound of high-intensity pixels and perform linear normalization. Other RGB images are also linearly normalized to the range (0,1) with 255 as the maximum value.

\section{Experiments}
\label{sec:eval}

As the first study to investigate image segmentation foundation models with federated learning, we conducted experiments to investigate three questions:

\noindent (1) FedFMS performance: Can SAM maintain its capabilities when trained under federated learning, comparable to SAM under centralized training?

\noindent (2) Model Efficiency: Is FedMSA a more efficient and cost-saving method in federated learning training? How does its performance compare to FedSAM?

\noindent (3) Pre-training Impact: Can pre-training on large dataset enrich the prior knowledge of our federated foundation models, thereby surpassing conventional ones?

\subsection{Experimental Settings}

We adopt two commonly used metrics, Dice (the Dice coefficient) and IOU (Intersection over Union), to quantitatively evaluate the segmentation results. 
We treat the dataset of each client as an unseen test set, while the data of each remaining client is divided into training and validation sets at a ratio of 9:1.

%\subsection{Implementation Details}
In the federated learning process, all clients use the same hyper-parameter settings, and the local model is trained using Adam optimizer with a batch size of 6. The momentum parameters for Adam are set to 0.9 and 0.999, respectively. The pre-trained model utilized is provided by SAM publicly. We conducted a total of 100 federated training rounds, with each local epoch set to 1. The framework is implemented using PyTorch and trained on NVIDIA A800. %Following the operations of the code provided by MSA, we trained adapters for the encoder and the whole decoder. 

\subsection{Results}
\subsubsection{Overall Comparison}
To explore the impact of federated learning on foundation model, we compare \model\ with MSA and compare \modelall\ with SAM. We use FedU-Net and FednnU-Net as baselines. U-Net~\cite{ronneberger2015Unet} is a commonly used and effective convolutional network for biomedical image segmentation. nnU-Net~\cite{isensee2021nnu} is a more robust method compared to U-Net.
We use the pre-trained ViT-B architecture of SAM as default. \model-L and SAM-L are extension experiments using ViT-L. Results are presented in Table \ref{tab:result} and Figure \ref{fig:result}.%We employed two variants of pre-train SAM (\ie, ViT-B and ViT-L), to train \model\ and MSA, denoted as \model-B, MSA-B, \model-L, and MSA-L, respectively.

% \noindent\textbf{Overall Performance Comparison}\\
\noindent \textbf{FedFMS performance} Both federated foundation models (\model, \modelall) and non-federated foundation models (MSA, SAM) achieve promising results across various tasks. 
\modelall, which fine-tunes all parameters, outperforms \model\ in Prostate, Nuclei, and Fundus segmentation. 
In Brain Tumor dataset, \model~outperforms \modelall. 
\model-L~and MSA-L with larger parameter count also perform similarly, and both outperform models based on Vit-B.
The results demonstrate the potential of SAM in federated learning for medical image segmentation.
This suggests the feasibility of further extending advanced federated learning algorithms to foundation models for the medical image domain. %, such as enhanced privacy protection, personalization, or alleviation of data distribution shifts,

The performance of \modelall, \model~and \model-L is significantly higher than FedU-Net and FednnU-Net. The preprocessing of nnU-Net under federated learning is limited in effectiveness, and its training on Non-IID datasets is also unstable.
Using pre-trained SAM is beneficial for medical image segmentation as it can mitigate unseen domain issues, attributed to its abundant background knowledge. The varying data quantities and distributions across clients result in inconsistent convergence directions among different clients. This inconsistency further leads to suboptimal performance of the globally aggregated model on the server. The foundation model demonstrates higher robustness and stability, which can alleviate the above issues. Therefore, through fine-tuning SAM, \model, \modelall~and \model-L can achieve advanced performance.% under the federated learning scenario.

\begin{table}[H]
\centering
\tiny
\setlength{\tabcolsep}{1pt}
\caption{The comparison of \modelall, SAM, \model, MSA, FedU-Net, and ablation variants (denoted as \textit{italics}) on different medical image datasets. The p-values between FedSAM and SAM, as well as between FedMSA and MSA, are both greater than 0.5.}
%\vspace{-5pt}
\begin{tabular}{|c|cccccccccccccc|}
\hline
Dataset                  & \multicolumn{14}{c|}{Prostate Cancer}                                                                                                                                                                                                                                                                                                                                                                                                                                                                                        \\ \hline
Client                   & \multicolumn{2}{c|}{BIDMC}                                                    & \multicolumn{2}{c|}{HK}                                                    & \multicolumn{2}{c|}{I2CVB}                                                    & \multicolumn{2}{c|}{ISBI}                                                    & \multicolumn{2}{c|}{ISBI1.5}                                                    & \multicolumn{2}{c|}{UCL}                                                    & \multicolumn{2}{c|}{Average}                         \\ \hline
Model & Dice & \multicolumn{1}{c|}{IOU} & Dice & \multicolumn{1}{c|}{IOU} & Dice & \multicolumn{1}{c|}{IOU} & Dice & \multicolumn{1}{c|}{IOU} & Dice & \multicolumn{1}{c|}{IOU} & Dice & \multicolumn{1}{c|}{IOU} & Dice & IOU\\ \hline
FedU-Net & 0.498 & \multicolumn{1}{c|}{0.498} & 0.684 & \multicolumn{1}{c|}{0.645} & 0.034 & \multicolumn{1}{c|}{0.023} & 0.649 & \multicolumn{1}{c|}{0.590} & 0.671 & \multicolumn{1}{c|}{0.640} & 0.563 & \multicolumn{1}{c|}{0.542} & 0.516 & 0.490\\ %\hline
FednnU-Net               & 0.457                               & \multicolumn{1}{c|}{0.375}          & 0.602                               & \multicolumn{1}{c|}{0.534}          & 0.448                               & \multicolumn{1}{c|}{0.372}          & 0.747                               & \multicolumn{1}{c|}{0.681}          & 0.612                               & \multicolumn{1}{c|}{0.544}          & 0.671                               & \multicolumn{1}{c|}{0.591}          & 0.590                               & 0.516          \\ \hline
\textbf{\modelall} & 0.810 & \multicolumn{1}{c|}{0.774} & 0.841 & \multicolumn{1}{c|}{0.808} & \textbf{0.798} & \multicolumn{1}{c|}{\textbf{0.768}} & 0.837 & \multicolumn{1}{c|}{0.792} & 0.785 & \multicolumn{1}{c|}{0.754} & 0.844 & \multicolumn{1}{c|}{0.809} & 0.819 & 0.784\\ %\hline
SAM & 0.793 & \multicolumn{1}{c|}{0.758} & 0.837 & \multicolumn{1}{c|}{0.799} & 0.731 & \multicolumn{1}{c|}{0.697} & 0.812 & \multicolumn{1}{c|}{0.759} & 0.786 & \multicolumn{1}{c|}{0.754} & 0.848 & \multicolumn{1}{c|}{0.807} & 0.801 & 0.762\\ %\hline
\textit{\modelall~(-PT)} & 0.688 & \multicolumn{1}{c|}{0.493} & 0.542 & \multicolumn{1}{c|}{0.490} & 0.583 & \multicolumn{1}{c|}{0.567} & 0.459 & \multicolumn{1}{c|}{0.393} & 0.378 & \multicolumn{1}{c|}{0.352} & 0.474 & \multicolumn{1}{c|}{0.449} & 0.521 & 0.457\\ %\hline
\textit{SAM~(-PT)} & 0.515 & \multicolumn{1}{c|}{0.500} & 0.679 & \multicolumn{1}{c|}{0.619} & 0.511 & \multicolumn{1}{c|}{0.486} & 0.577 & \multicolumn{1}{c|}{0.496} & 0.517 & \multicolumn{1}{c|}{0.470} & 0.520 & \multicolumn{1}{c|}{0.471} & 0.553 & 0.507\\ \hline
\textbf{\model} & 0.769 & \multicolumn{1}{c|}{0.737} & 0.809 & \multicolumn{1}{c|}{0.777} & 0.754 & \multicolumn{1}{c|}{0.720} & 0.821 & \multicolumn{1}{c|}{0.771} & 0.795 & \multicolumn{1}{c|}{0.765} & 0.859 & \multicolumn{1}{c|}{0.820} & 0.801 & 0.765\\ %\hline
MSA & 0.749 & \multicolumn{1}{c|}{0.711} & 0.813 & \multicolumn{1}{c|}{0.775} & 0.748 & \multicolumn{1}{c|}{0.716} & 0.803 & \multicolumn{1}{c|}{0.758} & 0.782 & \multicolumn{1}{c|}{0.752} & 0.841 & \multicolumn{1}{c|}{0.801} & 0.789 & 0.752\\ %\hline
\textit{\model~(-PT)} & 0.499 & \multicolumn{1}{c|}{0.499} & 0.527 & \multicolumn{1}{c|}{0.495} & 0.582 & \multicolumn{1}{c|}{0.564} & 0.565 & \multicolumn{1}{c|}{0.481} & 0.525 & \multicolumn{1}{c|}{0.504} & 0.419 & \multicolumn{1}{c|}{0.400} & 0.520 & 0.491\\ %\hline
\textit{MSA~(-PT)} & 0.546 & \multicolumn{1}{c|}{0.522} & 0.422 & \multicolumn{1}{c|}{0.389} & 0.474 & \multicolumn{1}{c|}{0.444} & 0.495 & \multicolumn{1}{c|}{0.423} & 0.476 & \multicolumn{1}{c|}{0.445} & 0.482 & \multicolumn{1}{c|}{0.427} & 0.482 & 0.442\\ \hline
\textbf{\model-L} & 0.806 & \multicolumn{1}{c|}{0.777} & \textbf{0.869} & \multicolumn{1}{c|}{\textbf{0.838}} & 0.772 & \multicolumn{1}{c|}{0.743} & \textbf{0.838} & \multicolumn{1}{c|}{\textbf{0.793}} & \textbf{0.821} & \multicolumn{1}{c|}{\textbf{0.793}} & 0.867 & \multicolumn{1}{c|}{\textbf{0.834}} & \textbf{0.829} & \textbf{0.796}\\ %\hline
MSA-L & \textbf{0.810} & \multicolumn{1}{c|}{\textbf{0.779}} & 0.845 & \multicolumn{1}{c|}{0.814} & 0.764 & \multicolumn{1}{c|}{0.742} & 0.836 & \multicolumn{1}{c|}{0.792} & 0.811 & \multicolumn{1}{c|}{0.782} & \textbf{0.869} & \multicolumn{1}{c|}{0.834} & 0.823 & 0.791\\ \hline
\end{tabular}
\smallskip

\begin{tabular}{|c|cccccccccc|c|cccc|}
\cline{1-11} \cline{13-16}
Dataset                  & \multicolumn{10}{c|}{Brain Tumor}                                                                                                                                                                                                                                                                                                                                    & ~ & \multicolumn{4}{c|}{Fundus}                                                                                                      \\ \cline{1-11} \cline{13-16} 
\multirow{2}{*}{Client}  & \multicolumn{2}{c|}{\multirow{2}{*}{1}}                                   & \multicolumn{2}{c|}{\multirow{2}{*}{6}}                                   & \multicolumn{2}{c|}{\multirow{2}{*}{18}}                                   & \multicolumn{2}{c|}{\multirow{2}{*}{21}}                                   & \multicolumn{2}{c|}{\multirow{2}{*}{Average}}        & ~ & \multicolumn{4}{c|}{REFUGE}                                                                                                           \\ \cline{13-16} 
                         & \multicolumn{2}{c|}{}                                                     & \multicolumn{2}{c|}{}                                                     & \multicolumn{2}{c|}{}                                                     & \multicolumn{2}{c|}{}                                                     & \multicolumn{2}{c|}{}                                & ~ & \multicolumn{2}{c|}{OD}                                                   & \multicolumn{2}{c|}{OC}                              \\  \cline{1-11} \cline{13-16} 
Model & Dice & \multicolumn{1}{c|}{IOU} & Dice & \multicolumn{1}{c|}{IOU} & Dice & \multicolumn{1}{c|}{IOU} & Dice & \multicolumn{1}{c|}{IOU} & Dice & IOU & ~ & Dice & \multicolumn{1}{c|}{IOU} & Dice & IOU \\ \cline{1-11} \cline{13-16}
FedU-Net & 0.860 & \multicolumn{1}{c|}{0.822} & 0.851 & \multicolumn{1}{c|}{0.806} & 0.861 & \multicolumn{1}{c|}{0.824} & 0.857 & \multicolumn{1}{c|}{0.817} & 0.857 & 0.817 & ~ & 0.848 & \multicolumn{1}{c|}{0.743} & 0.840 & 0.733 \\ %\cline{1-11} \cline{13-16}
FednnU-Net               & 0.772                               & \multicolumn{1}{c|}{0.711}          & 0.804                               & \multicolumn{1}{c|}{0.740}          & 0.781                               & \multicolumn{1}{c|}{0.721}          & 0.785                               & \multicolumn{1}{c|}{0.727}          & 0.786                               & 0.725          & ~ & 0.843                               & \multicolumn{1}{c|}{0.733}          & 0.796                               & 0.671          \\ \cline{1-11} \cline{13-16}
\textbf{\modelall} & 0.869 & \multicolumn{1}{c|}{0.831} & 0.880 & \multicolumn{1}{c|}{0.836} & 0.879 & \multicolumn{1}{c|}{0.839} & 0.860 & \multicolumn{1}{c|}{0.822} & 0.872 & 0.832 & ~ & 0.869 & \multicolumn{1}{c|}{0.772} & \textbf{0.873} & 0.781 \\ %\cline{1-11} \cline{13-16}
SAM & 0.867 & \multicolumn{1}{c|}{0.830} & 0.877 & \multicolumn{1}{c|}{0.833} & 0.876 & \multicolumn{1}{c|}{0.838} & 0.849 & \multicolumn{1}{c|}{0.809} & 0.867 & 0.828 & ~ & 0.859 & \multicolumn{1}{c|}{0.758} & 0.855 & 0.758 \\ %\cline{1-11} \cline{13-16}
\textit{\modelall~(-PT)} & 0.809 & \multicolumn{1}{c|}{0.767} & 0.853 & \multicolumn{1}{c|}{0.807} & 0.830 & \multicolumn{1}{c|}{0.790} & 0.832 & \multicolumn{1}{c|}{0.791} & 0.831 & 0.789 & ~ & 0.857 & \multicolumn{1}{c|}{0.756} & 0.842 & 0.736 \\ %\cline{1-11} \cline{13-16}
\textit{SAM~(-PT)} & 0.827 & \multicolumn{1}{c|}{0.785} & 0.851 & \multicolumn{1}{c|}{0.803} & 0.838 & \multicolumn{1}{c|}{0.798} & 0.818 & \multicolumn{1}{c|}{0.774} & 0.834 & 0.790 & ~ & 0.833 & \multicolumn{1}{c|}{0.721} & 0.821 & 0.710 \\ \cline{1-11} \cline{13-16}
\textbf{\model} & 0.877 & \multicolumn{1}{c|}{0.838} & 0.876 & \multicolumn{1}{c|}{0.832} & 0.884 & \multicolumn{1}{c|}{0.847} & 0.862 & \multicolumn{1}{c|}{0.823} & 0.875 & 0.835 & ~ & 0.860 & \multicolumn{1}{c|}{0.760} & 0.858 & 0.763 \\ %\cline{1-11} \cline{13-16}
MSA & 0.876 & \multicolumn{1}{c|}{0.837} & 0.871 & \multicolumn{1}{c|}{0.828} & 0.883 & \multicolumn{1}{c|}{0.845} & 0.853 & \multicolumn{1}{c|}{0.814} & 0.871 & 0.831 & ~ & \textbf{0.881} & \multicolumn{1}{c|}{0.792} & 0.866 & 0.773 \\ %\cline{1-11} \cline{13-16}
\textit{\model~(-PT)} & 0.808 & \multicolumn{1}{c|}{0.767} & 0.856 & \multicolumn{1}{c|}{0.811} & 0.811 & \multicolumn{1}{c|}{0.770} & 0.826 & \multicolumn{1}{c|}{0.784} & 0.825 & 0.783 & ~ & 0.846 & \multicolumn{1}{c|}{0.742} & 0.842 & 0.736 \\ %\cline{1-11} \cline{13-16}
\textit{MSA~(-PT)} & 0.837 & \multicolumn{1}{c|}{0.797} & 0.849 & \multicolumn{1}{c|}{0.804} & 0.844 & \multicolumn{1}{c|}{0.805} & 0.830 & \multicolumn{1}{c|}{0.790} & 0.840 & 0.799 & ~ & 0.845 & \multicolumn{1}{c|}{0.739} & 0.829 & 0.719 \\ \cline{1-11} \cline{13-16}
\textbf{\model-L} & \textbf{0.887} & \multicolumn{1}{c|}{\textbf{0.850}} & \textbf{0.883} & \multicolumn{1}{c|}{\textbf{0.841}} & \textbf{0.895} & \multicolumn{1}{c|}{\textbf{0.859}} & 0.876 & \multicolumn{1}{c|}{0.840} & \textbf{0.885} & \textbf{0.847} & ~ & 0.869 & \multicolumn{1}{c|}{0.772} & 0.869 & \textbf{0.867} \\ %\cline{1-11} \cline{13-16}
MSA-L & 0.886 & \multicolumn{1}{c|}{0.850} & 0.875 & \multicolumn{1}{c|}{0.832} & 0.890 & \multicolumn{1}{c|}{0.854} & \textbf{0.877} & \multicolumn{1}{c|}{\textbf{0.840}} & 0.882 & 0.844 & ~ & 0.879 & \multicolumn{1}{c|}{\textbf{0.794}} & 0.817 & 0.739 \\ \cline{1-11} \cline{13-16}
\end{tabular}

%\bigskip
\smallskip

\begin{tabular}{|c|cccccccccccccccc|}
\hline
Dataset                  & \multicolumn{16}{c|}{Fundus (Continued Table)}                                                                                                                                                                                                                                                                                                                                                                                                                                                                                                                                                           \\ \hline
\multirow{2}{*}{Client}   & \multicolumn{4}{c|}{ORIGA}                                                                                                                                & \multicolumn{4}{c|}{G1020}                                                                                                                                & \multicolumn{4}{c|}{Drishit-GS1}                                                                                                                                & \multicolumn{4}{c|}{Average}                                                                                                     \\ \cline{2-17}
                         & \multicolumn{2}{c|}{OD}                                                   & \multicolumn{2}{c|}{OC}                                                   & \multicolumn{2}{c|}{OD}                                                   & \multicolumn{2}{c|}{OC}                                                   & \multicolumn{2}{c|}{OD}                                                   & \multicolumn{2}{c|}{OC}                                                   & \multicolumn{2}{c|}{OD}                                                   & \multicolumn{2}{c|}{OC}                              \\ \hline 
Model    & \multicolumn{1}{c}{Dice}           & \multicolumn{1}{c|}{IOU}            & \multicolumn{1}{c}{Dice}           & \multicolumn{1}{c|}{IOU}            & \multicolumn{1}{c}{Dice}           & \multicolumn{1}{c|}{IOU}            & \multicolumn{1}{c}{Dice}           & \multicolumn{1}{c|}{IOU}            & \multicolumn{1}{c}{Dice}           & \multicolumn{1}{c|}{IOU}            & \multicolumn{1}{c}{Dice}           & \multicolumn{1}{c|}{IOU}            & \multicolumn{1}{c}{Dice}           & \multicolumn{1}{c|}{IOU}            & \multicolumn{1}{c}{Dice}           & IOU            \\ \hline
                         FedU-Net                 & 0.794          & \multicolumn{1}{c|}{0.667} & 0.830          & \multicolumn{1}{c|}{0.722} & 0.550          & \multicolumn{1}{c|}{0.427} & 0.449          & \multicolumn{1}{c|}{0.346} & 0.677          & \multicolumn{1}{c|}{0.529} & 0.779          & \multicolumn{1}{c|}{0.650} & 0.717          & \multicolumn{1}{c|}{0.592} & 0.725          & 0.613          \\ %\hline
                         FednnU-Net               & 0.771                               & \multicolumn{1}{c|}{0.636}          & 0.804                               & \multicolumn{1}{c|}{0.686}          & 0.541                               & \multicolumn{1}{c|}{0.404}          & 0.406                               & \multicolumn{1}{c|}{0.297}          & 0.679                               & \multicolumn{1}{c|}{0.536}          & 0.725                               & \multicolumn{1}{c|}{0.607}          & 0.709                               & \multicolumn{1}{c|}{0.577}          & 0.683                               & 0.565          \\ \hline
                         \textbf{\modelall}       & 0.816          & \multicolumn{1}{c|}{0.698} & 0.846          & \multicolumn{1}{c|}{0.748} & 0.717          & \multicolumn{1}{c|}{0.602} & 0.556          & \multicolumn{1}{c|}{0.456} & 0.715          & \multicolumn{1}{c|}{0.571} & 0.777          & \multicolumn{1}{c|}{0.643} & 0.779          & \multicolumn{1}{c|}{0.661} & 0.763          & 0.657          \\ %\hline
                         SAM                      & 0.820          & \multicolumn{1}{c|}{0.703} & 0.831          & \multicolumn{1}{c|}{0.729} & 0.621          & \multicolumn{1}{c|}{0.512} & 0.546          & \multicolumn{1}{c|}{0.444} & 0.676          & \multicolumn{1}{c|}{0.526} & 0.767          & \multicolumn{1}{c|}{0.632} & 0.744          & \multicolumn{1}{c|}{0.625} & 0.750          & 0.641          \\ %\hline
                         \textit{\modelall~(-PT)} & 0.765          & \multicolumn{1}{c|}{0.628} & 0.811          & \multicolumn{1}{c|}{0.697} & 0.420          & \multicolumn{1}{c|}{0.314} & 0.347          & \multicolumn{1}{c|}{0.266} & 0.537          & \multicolumn{1}{c|}{0.395} & 0.640          & \multicolumn{1}{c|}{0.490} & 0.645          & \multicolumn{1}{c|}{0.523} & 0.660          & 0.547          \\ %\hline
                         \textit{SAM~(-PT)}       & 0.777          & \multicolumn{1}{c|}{0.645} & 0.778          & \multicolumn{1}{c|}{0.777} & 0.482          & \multicolumn{1}{c|}{0.364} & 0.432          & \multicolumn{1}{c|}{0.335} & 0.509          & \multicolumn{1}{c|}{0.371} & 0.718          & \multicolumn{1}{c|}{0.579} & 0.650          & \multicolumn{1}{c|}{0.525} & 0.687          & 0.600          \\ \hline
                         \textbf{\model}          & 0.822          & \multicolumn{1}{c|}{0.705} & \textbf{0.850} & \multicolumn{1}{c|}{\textbf{0.753}} & 0.733          & \multicolumn{1}{c|}{0.614} & 0.560          & \multicolumn{1}{c|}{0.463} & 0.692          & \multicolumn{1}{c|}{0.545} & 0.769          & \multicolumn{1}{c|}{0.632} & 0.777          & \multicolumn{1}{c|}{0.656} & 0.759          & 0.653          \\ %\hline
                         MSA                      & 0.834          & \multicolumn{1}{c|}{0.723} & 0.844          & \multicolumn{1}{c|}{0.746} & 0.693          & \multicolumn{1}{c|}{0.573} & 0.551          & \multicolumn{1}{c|}{0.449} & 0.716          & \multicolumn{1}{c|}{\textbf{0.576}} & \textbf{0.798} & \multicolumn{1}{c|}{\textbf{0.672}} & 0.781          & \multicolumn{1}{c|}{0.666} & 0.765          & 0.660          \\ %\hline
                         \textit{\model~(-PT)}    & 0.789          & \multicolumn{1}{c|}{0.661} & 0.817          & \multicolumn{1}{c|}{0.704} & 0.475          & \multicolumn{1}{c|}{0.362} & 0.465          & \multicolumn{1}{c|}{0.374} & 0.557          & \multicolumn{1}{c|}{0.408} & 0.683          & \multicolumn{1}{c|}{0.536} & 0.667          & \multicolumn{1}{c|}{0.543} & 0.702          & 0.587          \\ %\hline
                         \textit{MSA~(-PT)}       & 0.804          & \multicolumn{1}{c|}{0.683} & 0.811          & \multicolumn{1}{c|}{0.704} & 0.573          & \multicolumn{1}{c|}{0.451} & 0.489          & \multicolumn{1}{c|}{0.393} & 0.481          & \multicolumn{1}{c|}{0.347} & 0.683          & \multicolumn{1}{c|}{0.536} & 0.676          & \multicolumn{1}{c|}{0.555} & 0.703          & 0.588          \\ \hline
                         \textbf{\model-L}        & 0.836          & \multicolumn{1}{c|}{0.725} & 0.845          & \multicolumn{1}{c|}{0.748} & 0.732          & \multicolumn{1}{c|}{0.630} & \textbf{0.587} & \multicolumn{1}{c|}{\textbf{0.494}} & 0.704          & \multicolumn{1}{c|}{0.559} & 0.762          & \multicolumn{1}{c|}{0.625} & 0.785          & \multicolumn{1}{c|}{0.672} & \textbf{0.766} & \textbf{0.684} \\ %\hline
                         MSA-L                    & \textbf{0.855} & \multicolumn{1}{c|}{\textbf{0.754}} & 0.846          & \multicolumn{1}{c|}{0.750} & \textbf{0.734} & \multicolumn{1}{c|}{\textbf{0.632}} & 0.582          & \multicolumn{1}{c|}{0.485} & \textbf{0.717} & \multicolumn{1}{c|}{0.573} & 0.776          & \multicolumn{1}{c|}{0.643} & \textbf{0.796} & \multicolumn{1}{c|}{\textbf{0.688}} & 0.755          & 0.654          \\ \hline
\end{tabular}
\smallskip

\begin{tabular}{|c|cccccccccccccccc|}
\hline
Dataset                  & \multicolumn{16}{c|}{Nuclei}                                                                                                                                                                                                                                                                                                                                                                                                                                                                                                                                                                             \\ \hline
Client                   & \multicolumn{2}{c|}{Adrenal}                                                    & \multicolumn{2}{c|}{Esophagus}                                                    & \multicolumn{2}{c|}{Bile-duct}                                                    & \multicolumn{2}{c|}{Uterus}                                                    & \multicolumn{2}{c|}{MoNuSAC}                                                    & \multicolumn{2}{c|}{TNBC}                                                    & \multicolumn{2}{c|}{MoNuSeg}                                                    & \multicolumn{2}{c|}{Average}                         \\ \hline
Model & Dice & \multicolumn{1}{c|}{IOU} & Dice & \multicolumn{1}{c|}{IOU} & Dice & \multicolumn{1}{c|}{IOU} & Dice & \multicolumn{1}{c|}{IOU} & Dice & \multicolumn{1}{c|}{IOU} & Dice & \multicolumn{1}{c|}{IOU} & Dice & \multicolumn{1}{c|}{IOU} & Dice & IOU \\ \hline
FedU-Net & 0.742 & \multicolumn{1}{c|}{0.618} & 0.781 & \multicolumn{1}{c|}{0.655} & 0.717 & \multicolumn{1}{c|}{0.598} & 0.777 & \multicolumn{1}{c|}{0.645} & 0.559 & \multicolumn{1}{c|}{0.404} & 0.703 & \multicolumn{1}{c|}{0.548} & 0.678 & \multicolumn{1}{c|}{0.519} & 0.708 & 0.569 \\ %\hline
FednnU-Net               & 0.798                               & \multicolumn{1}{c|}{0.684}          & 0.807                               & \multicolumn{1}{c|}{0.690}          & 0.754                               & \multicolumn{1}{c|}{0.642}          & 0.803                               & \multicolumn{1}{c|}{0.681}          & 0.588                               & \multicolumn{1}{c|}{0.435}          & 0.747                               & \multicolumn{1}{c|}{0.606}          & 0.713                               & \multicolumn{1}{c|}{0.563}          & 0.744                               & 0.614          \\ \hline
\textbf{\modelall} & 0.810 & \multicolumn{1}{c|}{0.698} & 0.807 & \multicolumn{1}{c|}{0.693} & 0.765 & \multicolumn{1}{c|}{0.659} & \textbf{0.832} & \multicolumn{1}{c|}{\textbf{0.717}} & 0.623 & \multicolumn{1}{c|}{0.472} & \textbf{0.776} & \multicolumn{1}{c|}{\textbf{0.643}} & 0.746 & \multicolumn{1}{c|}{0.602} & 0.765 & 0.640 \\ %\hline
SAM & 0.819 & \multicolumn{1}{c|}{0.709} & 0.777 & \multicolumn{1}{c|}{0.655} & 0.768 & \multicolumn{1}{c|}{0.665} & 0.824 & \multicolumn{1}{c|}{0.706} & 0.606 & \multicolumn{1}{c|}{0.457} & 0.709 & \multicolumn{1}{c|}{0.564} & 0.677 & \multicolumn{1}{c|}{0.518} & 0.740 & 0.611 \\ %\hline
\textit{\modelall~(-PT)} & 0.666 & \multicolumn{1}{c|}{0.533} & 0.736 & \multicolumn{1}{c|}{0.597} & 0.690 & \multicolumn{1}{c|}{0.561} & 0.742 & \multicolumn{1}{c|}{0.598} & 0.572 & \multicolumn{1}{c|}{0.415} & 0.636 & \multicolumn{1}{c|}{0.485} & 0.629 & \multicolumn{1}{c|}{0.468} & 0.667 & 0.523 \\ %\hline
\textit{SAM~(-PT)} & 0.686 & \multicolumn{1}{c|}{0.548} & 0.711 & \multicolumn{1}{c|}{0.571} & 0.669 & \multicolumn{1}{c|}{0.539} & 0.713 & \multicolumn{1}{c|}{0.564} & 0.587 & \multicolumn{1}{c|}{0.434} & 0.667 & \multicolumn{1}{c|}{0.513} & 0.643 & \multicolumn{1}{c|}{0.479} & 0.668 & 0.521 \\ \hline
\textbf{\model} & 0.806 & \multicolumn{1}{c|}{0.694} & 0.802 & \multicolumn{1}{c|}{0.688} & 0.763 & \multicolumn{1}{c|}{0.655} & 0.805 & \multicolumn{1}{c|}{0.682} & 0.640 & \multicolumn{1}{c|}{0.490} & 0.730 & \multicolumn{1}{c|}{0.600} & 0.741 & \multicolumn{1}{c|}{0.598} & 0.755 & 0.630 \\ %\hline
MSA & 0.813 & \multicolumn{1}{c|}{0.702} & 0.805 & \multicolumn{1}{c|}{0.693} & 0.769 & \multicolumn{1}{c|}{0.664} & 0.824 & \multicolumn{1}{c|}{0.707} & 0.629 & \multicolumn{1}{c|}{0.477} & 0.630 & \multicolumn{1}{c|}{0.486} & 0.665 & \multicolumn{1}{c|}{0.506} & 0.733 & 0.605 \\ %\hline
\textit{\model~(-PT)} & 0.692 & \multicolumn{1}{c|}{0.558} & 0.728 & \multicolumn{1}{c|}{0.589} & 0.684 & \multicolumn{1}{c|}{0.556} & 0.739 & \multicolumn{1}{c|}{0.596} & 0.562 & \multicolumn{1}{c|}{0.406} & 0.628 & \multicolumn{1}{c|}{0.477} & 0.642 & \multicolumn{1}{c|}{0.482} & 0.668 & 0.523 \\ %\hline
\textit{MSA~(-PT)} & 0.649 & \multicolumn{1}{c|}{0.507} & 0.662 & \multicolumn{1}{c|}{0.523} & 0.696 & \multicolumn{1}{c|}{0.552} & 0.662 & \multicolumn{1}{c|}{0.527} & 0.559 & \multicolumn{1}{c|}{0.406} & 0.622 & \multicolumn{1}{c|}{0.458} & 0.658 & \multicolumn{1}{c|}{0.517} & 0.644 & 0.499 \\ \hline
\textbf{\model-L} & 0.811 & \multicolumn{1}{c|}{0.701} & 0.805 & \multicolumn{1}{c|}{0.699} & 0.767 & \multicolumn{1}{c|}{0.662} & 0.824 & \multicolumn{1}{c|}{0.706} & \textbf{0.644} & \multicolumn{1}{c|}{\textbf{0.494}} & 0.769 & \multicolumn{1}{c|}{0.635} & \textbf{0.761} & \multicolumn{1}{c|}{\textbf{0.622}} & \textbf{0.769} & \textbf{0.646} \\ %\hline
MSA-L & \textbf{0.820} & \multicolumn{1}{c|}{\textbf{0.711}} & \textbf{0.814} & \multicolumn{1}{c|}{\textbf{0.703}} & \textbf{0.802} & \multicolumn{1}{c|}{\textbf{0.703}} & 0.823 & \multicolumn{1}{c|}{0.704} & 0.630 & \multicolumn{1}{c|}{0.480} & 0.683 & \multicolumn{1}{c|}{0.542} & 0.700 & \multicolumn{1}{c|}{0.547} & 0.753 & 0.627 \\ \hline
\end{tabular}
%\vspace{-0.1in}
\label{tab:result}
\end{table}

\noindent\textbf{Further Discussion} 
The test results of \model~and MSA are generally similar across various datasets, and the performance of \modelall~and SAM is also similar. The federated learning paradigm leads to slight differences in their performance. The discrepancies are slightly larger in their tests on clients TNBC and MoNuSeg for Nuclei segmentation. The Nuclei dataset has the smallest dataset among all the datasets. 
Moreover, there are large differences among different types of cells. These factors lead to inconsistent convergence directions in federated learning. % In federated learning, the small data size for each client may result in distinct convergence outcomes for the models. %, as well as on clients A and D for Lung segmentation

\subsubsection{Model Efficiency Analysis} 
We calculate the learnable parameter count, training time in GPU (the average training time in Fundus dataset), GPU memory usage in each client, and the estimated FLOPs (Floating Point Operations) for both forward and backward propagation for \model~and \modelall, as shown in Table \ref{tab:efficiency}. The number of parameters (denoted as $n$) to be trained and updated determines the model's training speed and communication cost. The amount of parameters to be communicated in each round of federated learning is $2n$. The results show that \model~freezes a large number of parameters in the encoder, resulting in a significantly reduced parameter count and FLOPs compared to \modelall, and consequently reducing communication and training costs. We calculate the average time required to predict each 2D image, as shown in "Predicting time" in Table 2. During the prediction process, FedMSA takes a bit longer because it has more parameters from adapters compared to FedSAM. %During the prediction process, FedMSA takes a bit longer because it has more parameters from adapters compared to FedSAM.
\begin{table}[t]
\centering
% \ssmall
\scriptsize
 %\footnotesize
% \small
%\tiny
\setlength{\tabcolsep}{0.5pt}
%\vspace{-0.1in}
\caption{Model efficiency analysis on \model~and \modelall.}
%\vspace{-5pt}
\iffalse
\begin{tabular}{c|cc|cc|cc|cc}
\hline
Index & \multicolumn{2}{c|}{Learnable Parameter} & \multicolumn{2}{c|}{Average Training Time} & \multicolumn{2}{c|}{GPU Memory Usage}        & \multicolumn{2}{c}{FLOPs}              \\ \hline
Model & \multicolumn{1}{c|}{\makebox[4.5em]{\model}}  & \modelall & \multicolumn{1}{c|}{\makebox[5em]{\model}}    & \modelall & \multicolumn{1}{c|}{\model}     & \modelall  & \multicolumn{1}{c|}{\model} & \modelall \\ \hline
Value & \multicolumn{1}{c|}{14.7 B}  & 93.7 B    & \multicolumn{1}{c|}{739.9 min} & 911.4 min & \multicolumn{1}{c|}{52,274 MiB} & 58,478 MiB & \multicolumn{1}{c|}{5.7 T}  & 13.4 T    \\ \hline
\end{tabular}
\fi
\begin{tabular}{cc|cc|cc|cc|cc}
\hline
\multicolumn{2}{c|}{Learnable Parameter} & \multicolumn{2}{c|}{Training Time} & \multicolumn{2}{c|}{GPU Memory Usage}        & \multicolumn{2}{c|}{FLOPs}              & \multicolumn{2}{c}{Predicting Time}      \\ \hline
\multicolumn{1}{c|}{\model}  & \modelall & \multicolumn{1}{c|}{\model}    & \modelall & \multicolumn{1}{c|}{\model}     & \modelall  & \multicolumn{1}{c|}{\model} & \modelall & \multicolumn{1}{c|}{\model}  & \modelall \\ \hline
\multicolumn{1}{c|}{14.7 B}  & 93.7 B    & \multicolumn{1}{c|}{739.9 min} & 911.4 min & \multicolumn{1}{c|}{52,274 MiB} & 58,478 MiB & \multicolumn{1}{c|}{5.7 T}  & 13.4 T    & \multicolumn{1}{c|}{0.127 s} & 0.118 s   \\ \hline
\end{tabular}
\label{tab:efficiency}
%\vspace{-5pt}
\end{table}
\subsubsection{Pre-training Impact}
SAM is pre-trained on a large-scale natural dataset. To investigate the effectiveness of this pretraining for medical image segmentation and its impact on federated learning, we conducted an ablation study. For \modelall, \model, SAM and MSA, we constructed variants \textit{\modelall~(-PT)}, \textit{\model~(-PT)}, \textit{SAM~(-PT)} and \textit{MSA~(-PT)} without using pre-trained parameters, and the experimental results are illustrated in Table \ref{tab:result}. 
The results show that not using pre-trained parameters from SAM leads to a drastic decrease in performance. 
In some cases, \textit{\modelall~(-PT)} performs worse than \textit{SAM (-PT)}, which is due to the convergence of no pre-trained SAM under federated learning is not stable. % which is due to the inconsistent convergence directions of different clients in federated learning. 
\textit{\modelall~(-PT)} and \textit{\model~(-PT)} sometimes perform worse than the lightweight model FedU-Net, for example in experiments Adrenal, Esophagus, Bile-duct, Uterus, TNBC, MoNuSeg on Nuclei dataset and HK, ISBI, ISBI1.5, UCL on Prostate Cancer dataset. This indicates that the pretraining knowledge of SAM is crucial for its effectiveness under the federated learning paradigm, which enables our federated foundation models to far surpass the traditional federated learning models (\eg, FednnU-Net). The code for our implementation of FednnU-Net is available at https://github.com/LMIAPC/FednnU-Net. 

\section{Conclusion}
\label{sec:conlusion}
In this study, we propose a solution to deploy the foundation model SAM and its efficient variant MSA within the federated learning framework, referred to as \modelall~and \model~respectively. We collected various multi-institutional federated datasets for our experiment.
By leveraging rich pre-training knowledge, \modelall~and \model~demonstrate excellent performance in addressing the inherent training issues of federated learning, achieving comparable results to the foundation models in centralized training. Additionally, we conducted an efficiency analysis between \model~and \modelall.
Our study is the first to introduce foundation models for federated learning in the medical image domain. It will encourage the integration of more foundation models into privacy-preserving federated learning frameworks, which holds profound practical significance.

\begin{credits}
\subsubsection{\ackname} This work is supported by Shenzhen Science and Technology Program (No.JCYJ20230807120800001), and 2023 Shenzhen sustainable supporting funds for colleges and universities (No.20231121165240001). The authors sincerely appreciate the computing environment supported by the China Unicom Shenzhen Intelligent Computing Center.

\subsubsection{\discintname}
The authors have no competing interests to declare that are
relevant to the content of this article.
\end{credits}

\bibliographystyle{splncs04}
\bibliography{mybibliography}

\end{document}